\documentclass[aps,prb,amssymb,12pt,notitlepage,article]{revtex4-1}
\usepackage{graphicx}
\usepackage{amsmath}
\usepackage{color}
\usepackage{enumitem}
\usepackage{graphicx}
\usepackage[normalem]{ulem}
\usepackage{comment}

\newif\ifHighlitedChanges
\def\ifHighlitedChanges{\iffalse}
\ifHighlitedChanges
  
  \def\STRIKE#1{{\color{red}\sout{#1}}}
\else
  
  \def\STRIKE#1{\relax}
\fi
\begin{document}

\bibliographystyle{apsrev4-1-titles}
\title{Analytical solutions of the Arrhenius-Semenov problem \\ for constant volume burn}
\author{Galen T. Craven}  
%\email{}
\affiliation{Theoretical Division, Los Alamos National Laboratory, Los Alamos, New Mexico, USA} 

\begin{abstract}
%The Semenov problem for describes. Under
%this reduces to solving an single ordinary differential equation that describes the progress of the
%of the reaction progress variable. Here, 
Analytical solutions to the Semenov thermal ignition problem for constant volume burn governed by Arrhenius reaction kinetics are derived.
Specifically, an approximate analytical solution technique for the Arrhenius-Semenov differential equation is derived for reaction orders $n \in \mathbb{R}_{>0}$ 
and exact solutions are also constructed for reaction orders $n \in \mathbb{N}: n \leq 3$.
The approximation technique relies 
on expansion of the respective nondominant terms in the differential equation at the lower and upper bounds of the reaction progress variable
in order to create a pair of integrable series.
The two integrated series are then connected to create a single continuous analytical solution.
%Analysis of the developed approximation is presented,
%with excellent
Excellent agreement is observed between the analytical approximation and solutions obtained numerically.
The presented approximation constitutes a simple and robust strategy for solving the Arrhenius-Semenov problem analytically.
%These results physical understanding of constant burn and computational
%numerically solve this system.
\end{abstract}
%\pacs{82.20.Db, 05.40.Ca, 05.45.-a, 34.10.+x}
 \maketitle
\section{Introduction  \label{sec:Introduction}}

Theories describing equilibrium\cite{Uzer02,Pollak2005,dawn05a} and nonequilibrium\cite{Bazant2013nonequilibrium,craven15c,craven16c,matyushov16c,craven16a,craven14c,Santa2020} chemical kinetics 
play a fundamental role in 
%theoretical and
 computational treatments of shock-induced chemical reactions and high explosive detonations \cite{Tarver2007nonequilibrium,Menikoff2009,Kunova2016,Aslam2018AWSD,Menikoff2015,Menikoff2019,Peterson2020,Huber2020}. The principal motivations for developing these theories are to (a) determine chemical reaction rates, (b) to predict reaction products, and (c) to elucidate complex molecular behavior that arises due to 
%intra- and inter-
energy and charge redistribution that occurs as the reaction proceeds. While various advanced computational methods \cite{Ceriotti2021review, butler2018machine} 
can provide high-level insight into a system's 
%chemical dynamics, 
properties \cite{Houston2019,Valleau2020,Greaves2021, Liu2019,craven20c}, 
%both static and dynamic,
it is often advantageous to describe 
%reactions 
molecular processes
using analytical methods, for example, to develop physical intuition about the system or for computational efficiency. In complex systems, however, nonlinearities typically preclude exact analytical solutions to a system's dynamical equations and, therefore, approximation methods must be employed. In this report, we develop an analytical approximation method to solve the kinetic equation describing a constant volume burn governed by Arrhenius reaction kinetics.
%%%

Constant volume burn occurs when a thermal explosion takes place in a rigid reaction vessel over a fast enough time scale   
such that no heat is exchanged across the vessel boundary and the chemical and thermodynamic properties of the system are spatially uniform \cite{Menikoff2017,Boddington1977}.
Following Menikoff's mathematical exposition for constant volume burn from Ref.~\citenum{Menikoff2017},
%constant volume burn occurs when a thermal explosion takes place in a rigid reaction vessel over a fast enough time scale   
%such that no heat is exchanged across the vessel boundary and the chemical and thermodynamic properties of the system are spatially uniform. 
the ordinary differential equation (ODE) that describes the temperature evolution of a constant volume burn is \cite{Menikoff2017}
\begin{equation}
\label{eq:ASsys}
%\frac{d \lambda}{d t} &= \mathcal{R}(\lambda,T), 
%\\[1ex]
\frac{d T}{d t} = \frac{\mathcal{Q}}{C_v(\lambda,T)}\mathcal{R}(\lambda,T),
\end{equation}
where $\lambda \in \left[0,1\right]$ is a reaction progress variable,  $\mathcal{R}$ is the reaction rate, 
$\mathcal{Q}$ is the specific energy released, and $C_v$ is the specific heat.
When $\lambda = 0$ the system consists of completely unburned reactant and when $\lambda = 1$ the reactant is completely burned.
The initial ($\lambda = 0$) system temperature  is $T_0$ and the final ($\lambda = 1$) system temperature at the end of the reaction is $T_1$. 
Equation~(\ref{eq:ASsys}) is coupled with an equation that describes the time evolution of the reaction progress variable
\begin{equation}
\frac{d \lambda}{d t} = \mathcal{R}(\lambda,T).
\end{equation}
Here, the reaction rate is assumed to take the Arrhenius form
\begin{equation}
\mathcal{R}(\lambda,T) = (1-\lambda)^n k \exp\left[-\frac{T_a}{T(\lambda)}\right],
\end{equation}
where $n$ is the reaction order, $k$ is a pre-exponential frequency factor, and
$T_a$ is the activation temperature.
Assuming that $\mathcal{Q}$ and $C_v$ are constant,
the temperature of the system as a function of reaction progress is
\begin{equation}
\label{eq:T}
T(\lambda) = T_0 + (\mathcal{Q}/C_v) \lambda = T_0 + (T_1-T_0) \lambda,
\end{equation}
with $T_1 = T_0 + \mathcal{Q}/C_v$.
The time-dependence of $T$ in Eq.~(\ref{eq:T}) arises through the reaction progress variable $\lambda \rightarrow \lambda(t)$.
Under these conditions, the time evolution of the reaction can be described by the single ODE 
\begin{equation}
\label{eq:AS}
\frac{d \lambda}{d t} = (1-\lambda)^n k \exp\left[-\frac{T_a}{T_0 + (\mathcal{Q}/C_v) \lambda}\right] = (1-\lambda)^n k \exp\left[-\frac{T_a}{T_0 + (T_1-T_0)\lambda}\right].
\end{equation}
%which expresses the time evolution of the progress variable.
% through the reaction rate.

%which is the Arrhenius-Semenov (AS) equation for constant volume burn.
Our aim is to develop analytical techniques that can be used to solve Eq.~\eqref{eq:AS} over a broad class of reaction orders.
Note that this ODE is separable,
\begin{equation}
\label{eq:separable}
\int k dt = \int (1-\lambda)^{-n} \exp\left[\frac{T_a}{T_0 + (T_1-T_0) \lambda}\right] d \lambda,
\end{equation}
and, therefore, the Semenov ignition problem in the presence of Arrhenius rate kinetics can in essence be solved by evaluating the integral on the RHS of Eq.~(\ref{eq:separable}):
\begin{equation}
\label{eq:In}
I_n (\lambda) \equiv \int (1-\lambda)^{-n} \exp\left[\frac{T_a}{T_0 + (T_1-T_0) \lambda}\right] d \lambda.
\end{equation}

Below, in Sec.~\ref{sec:approx} we illustrate an analytical technique for approximating $I_n$ 
and, therefore, for solving the Semenov problem. In Sec.~\ref{sec:exact} we derive exact solutions for reaction orders $n \in \mathbb{N}: n \leq 3$.
Section~\ref{sec:results} contains a comparison between the developed approximation and numerical solutions for various reaction conditions.

%\CR{Talk about reaction orders use Menikoff paper}

%\section{Approximate Solution Through Series Expansion for $n \in \mathbb{R}^+$}

\section{Approximate Solution for $n \in \mathbb{R}_{>0}$\label{sec:approx}}

An exact solution of $I_n$ is not known for $n \in \mathbb{R}^{+} \setminus \mathbb{N}$.
Below we give an approximate analytical solution for $n \in \mathbb{R}_{>0}$ which relies on 
series expansion of different terms in the integrand 
at different ends of the $\lambda \in \left[0,1\right]$ interval. 
%For simplicity, all constants of integration are suppressed throughout this Section 
%except in the final solution.
%The reaction progress variable ranges from $0 $ to $1$.

%%\CR{General Mathematics}
%%
%%Consider the integral
%%
%%\begin{equation}
%%\label{eq:Inint}
%% \int f(\lambda) g(\lambda) d \lambda.
%%\end{equation}
%%where the functions $f$ and $g$ are related at the limits
%%\begin{equation}
%%\lim_{\lambda\to b}  \frac{f(\lambda)}{g(\lambda)} \ll 1.
%%\end{equation} 
%%and
%%\begin{equation}
%%\lim_{\lambda\to a}  \frac{g(\lambda)}{f(\lambda)}  \gg 1.
%%\end{equation} 
%%
%%for a where the support of $f$ and $g$ is $\lambda \in \left[a,b\right]$.
%%
%%\begin{equation}
%%\label{eq:Inint}
%% \int f(\lambda) (g_1(\lambda)+g_1(\lambda)+g_1(\lambda)) d \lambda + \begin{equation}
%%\label{eq:Inint}
%% \int S(f(\lambda)) g(\lambda) d \lambda 
%%\end{equation}
%%\end{equation}

As $\lambda \to 1$, the term $(1-\lambda)^{-n}$ dominates 
the integrand for $n>0$: 
\begin{equation}
\lim_{\lambda\to 1}  \frac{\exp\left[T_a / T(\lambda) \right]}{(1-\lambda)^{-n}} = 0.
\end{equation}
In this fully-burned limit, we therefore expand $\exp\left[T_a / T(\lambda)\right]$ about $\lambda = 1$
to obtain the approximation
\begin{equation}
\int (1-\lambda)^{-n} \exp\left[\frac{T_a}{T(\lambda)}\right] d \lambda \approx \int \sum^\alpha_{m = 0} (1-\lambda)^{m-n} \frac{D^m_\lambda \exp\left[\frac{T_a}{T(\lambda)}\right]\bigg|_{\lambda = 1}}{m!} \, d \lambda,
\end{equation}
where $\alpha$ is the series expansion order and $D^m_\lambda$ is $m$-th order derivative operator with respect to $\lambda$.
The integrals in the series approximation up to third order ($\alpha = 3$) are
\begin{equation}
\begin{aligned}
\label{eq:fullpos}
I^+_n (\lambda) &\approx \underbrace{\int (1-\lambda)^{-n} \exp\left[\frac{T_a}{T_1}\right]  d \lambda}_{I^{(0+)}_n} \\[1ex]
& + \underbrace{\int (1-\lambda)^{1-n} \exp\left[\frac{T_a}{T_1}\right] \left(\frac{\mathcal{Q}}{C_v}\right)\frac{T_a}{T^2_1} \, d \lambda}_{I^{(1+)}_n}\\[1ex] 
& + \underbrace{\int (1-\lambda)^{2-n} \exp\left[\frac{T_a}{T_1}\right]\left(\frac{\mathcal{Q}}{C_v}\right)^2\frac{T_a (T_a+2 T_1)}{2 T^4_1} \, d \lambda}_{I^{(2+)}_n}\\[1ex]
& + \underbrace{\int (1-\lambda)^{3-n} \exp\left[\frac{T_a}{T_1}\right] \left(\frac{\mathcal{Q}}{C_v}\right)^3 \frac{T_a (T^2_a+ 6 T_1^2+6 T_a T_1)}{6 T^6_1}\, d \lambda}_{I^{(3+)}_n},
\end{aligned}
\end{equation}
where the superscript ``$+$'' in $I^+_n$ denotes that the approximate solution is approached from the upper bound of $\lambda$
and $I^{(j+)}_n$ denotes the $j$th order term in the upper-bound expansion. 
%The general formula is
%\begin{equation}
%\mathcal{I}^{(\alpha-)}_n &= \int (1-\lambda)^{\alpha-n} \exp\left[\frac{T_a}{T_1}\right] \left(\frac{\mathcal{Q}}{C_v}\right)^\alpha 
%\frac{T_1^{-2 \alpha}}{\alpha!} \sum^{\alpha}_{q=1} T^q_a T^{\alpha-q}_1 d \lambda
%\end{equation}
Evaluating each integral in the series yields 
\begin{align}
I^{(0+)}_n &= \frac{(1-\lambda)^{1-n}}{n-1} \exp\left[\frac{T_a}{T_1}\right] \label{eq:intinit}, \\[1ex]
I^{(1+)}_n &= \frac{(1-\lambda)^{2-n}}{n-2} \exp\left[\frac{T_a}{T_1}\right] \left(\frac{\mathcal{Q}}{C_v}\right) \frac{T_a}{T^2_1},\\[1ex]
I^{(2+)}_n &= \frac{(1-\lambda)^{3-n}}{n-3} \exp\left[\frac{T_a}{T_1}\right] \left(\frac{\mathcal{Q}}{C_v}\right)^2\frac{T_a (T_a+2 T_1)}{2 T^4_1},\\[1ex]
I^{(3+)}_n &= \frac{(1-\lambda)^{4-n}}{n-4} \exp\left[\frac{T_a}{T_1}\right] \left(\frac{\mathcal{Q}}{C_v}\right)^3\frac{T_a (T^2_a +  6 T^2_1 + 6 T_a T_1)}{6 T^6_1}\label{eq:intfinal}.
\end{align}
%%%the general formula
%%%\begin{equation}
%%%\mathcal{I}^{(\alpha-)}_n &= \frac{(1-\lambda)^{\alpha+1-n}}{n-\alpha -1} \exp\left[\frac{T_a}{T_1}\right] \left(\frac{\mathcal{Q}}{C_v}\right)^\alpha 
%%%T_1^{-2 \alpha} \sum^{\alpha}_{q=1} T^q_a T^{\alpha-q}_1
%%%\end{equation}
Note that each of the $I^{(j+)}_n$ functions above is vertically asymptotic for $n=j+1$. However, as illustrated below in Sec.~\ref{sec:approxsub}, after applying the initial value to the full solution, there is no asymptote with respect to variation of $n$.

In the limit $\lambda \to 0$, the ratio between the two functions in the integrand of $I_n$ is
\begin{equation}
\lim_{\lambda\to 0} \frac{\exp\left[T_a / T(\lambda) \right]}{(1-\lambda)^{-n}} = \exp\left[\frac{T_a}{T_0}\right]. 
\end{equation} 
In this limit, 
we approximate the integral by expanding the $(1-\lambda)^{-n}$ term about $\lambda = 0$:
\begin{equation}
\label{eq:approx0}
\int (1-\lambda)^{-n} \exp\left[\frac{T_a}{T(\lambda)}\right] d \lambda \approx \int \sum^\alpha_{m = 0} \exp\left[\frac{T_a}{T(\lambda)}\right] \lambda^m \frac{D^m_\lambda (1-\lambda)^{-n} \bigg|_{\lambda = 0}}{m!} \, d \lambda.
\end{equation}
Expanding the RHS of Eq.~(\ref{eq:approx0}) up to third order gives the following approximation for $I_n$:
\begin{equation}
\begin{aligned}
I^-_n(\lambda) &\approx \underbrace{\int \exp\left[\frac{T_a}{T(\lambda)}\right] d \lambda}_{I^{(0-)}_n} \\[1ex] 
&+ \underbrace{\int \exp\left[\frac{T_a}{T(\lambda)}\right]  n \lambda \, d \lambda}_{I^{(1-)}_n}\\[1ex] 
&+ \underbrace{\int \exp\left[\frac{T_a}{T(\lambda)}\right] \frac{(n+n^2)}{2}  \lambda^2 \, d \lambda}_{I^{(2-)}_n} \\[1ex] 
&+ \underbrace{\int \exp\left[\frac{T_a}{T(\lambda)}\right] \frac{(2 n+3 n^2+n^3)}{6} \lambda^3 \, d \lambda}_{I^{(3-)}_n},
\end{aligned}
\end{equation}
where the superscript ``$-$'' denotes that the approximate solution is approached from the lower bound of $\lambda$
and $I^{(j-)}_n$ denotes the $j$th-order term in the lower-bound expansion. 

%%%%%%%%%%%%%%\CR{Math is good to here}
%%%%%%%%%%%%%\CR{Have checked the first two}
Evaluating these integrals gives 
\begin{align}
I^{(0-)}_n &=  \left(\frac{C_v}{\mathcal{Q}}\right) \Bigg[ T(\lambda)\exp\Big[h(\lambda)\Big] - T_a\text{Ei}\Big[h(\lambda)\Big] \Bigg],%\\[1ex]
\end{align}
%%%%%%%%%%%%%%%%%%%%%%%%%%%%%%%%%%%%%%%%%%%%%%
\begin{align}
I^{(1-)}_n &= \frac{n}{2} \left(\frac{C_v}{\mathcal{Q}}\right)^2 \Bigg[T(\lambda)\left(\frac{\mathcal{Q}}{C_v} \lambda - T_0+T_a\right)\exp\Big[h(\lambda)\Big]+ T_a(2 T_0-T_a) \text{Ei}\Big[h(\lambda)\Big]\Bigg], %\\[1ex]
\end{align}
%%%%%%%%%%%%%%%%%%%%%%%%%%%%%%%%%%%%%%%%%%%%%%
\begin{align}
\nonumber I^{(2-)}_n &= \frac{n+ n^2}{12} \left(\frac{C_v}{\mathcal{Q}}\right)^3 \\[1ex]
&\nonumber \quad \times 
 \Bigg[ T(\lambda) \left( 2 \left(\frac{\mathcal{Q}}{C_v}\right)^2 \lambda^2 + \left(\frac{\mathcal{Q}}{C_v} T_a - 2 T_0 T_1 + 2 T_0^2 \right) \lambda  
+ \left(T_a^2 - 5 T_a T_0 + 2 T_0^2\right) \right)   \\[1ex] 
& \quad \times \exp\Big[h(\lambda)\Big] - T_a \left(T_a^2 - 6 T_a T_0 + 6 T_0^2\right)\text{Ei}\Big[h(\lambda)\Big] \Bigg],%\\[1ex]
\end{align}
%%%%%%%%%%%%%%%%%%%%%%%%%%%%%%%%%%%%%%%%%%%%%%%%
\begin{align}
\nonumber I^{(3-)}_n &= \frac{(2 n+3 n^2+n^3)}{144} \left(\frac{C_v}{\mathcal{Q}}\right)^4 \Bigg[ \Bigg( 
6  \left(\frac{\mathcal{Q}}{C_v}\right)^4 \lambda^4  + 2 T_a  \left(\frac{\mathcal{Q}}{C_v}\right)^3 \lambda^3
- \left(\frac{\mathcal{Q}}{C_v}\right)^2 T_a (6 T_0 - T_a) \lambda^2 \\[1ex]
& \nonumber \quad + \left(\frac{\mathcal{Q}}{C_v}\right) T_a (T_a^2 - 10 T_a T_0 + 18 T_0^2) \lambda
- T_0 (3 T_0 - T_a)(T_a^2 -8 T_a T_0 + 2 T_0^2) \Bigg) \exp\Big[h(\lambda)\Big] \\[1ex]
&\quad - T_a(T_a^3 - 12 T_a^2 T_0 + 36 T_a T_0^2 - 24 T_0^3) \text{Ei}\Big[h(\lambda)\Big] \Bigg].
\end{align}
where
\begin{equation}
\label{eq:h}
h(\lambda) = \frac{T_a}{T(\lambda)}.
\end{equation}
%Terms for expansion orders greater than $\alpha = 3$ can also be added with after significant algebraic manipulation.

We now enforce the condition that the complete solution must be continuous in $t$. 
First, we apply the inital value $\lambda(0) = 0$ to both solution branches leading to the relations
\begin{equation}
k t^+(\lambda) = I^+_n(\lambda) - I^+_n(0),
\end{equation}
for the upper-bound approximation and
\begin{equation}
k t^-(\lambda) = I^-_n(\lambda) - I^-_n(0),
\end{equation}
for the lower-bound approximation, where $t^\pm (\lambda)$ denotes time as a function of $\lambda$ given by the respective ``$\pm$'' solution.
%For large $t$ (the exact solution to $I_n$), $t \approx t^+$, and for small $t$, $t \approx t^-$. In general, however, $t^+ \neq t^-$.
%Solutions to the ODE are
We want to enforce the condition $t^+(\lambda^*) = t^-(\lambda^*)$ at some point $\lambda^*$ in the interval $\left[0,1\right]$
and then to connect the two series using a continuous piecewise function that shifts between branches at $\lambda^*$.
The method we apply to construct this continuous function is to shift the upper-bound approximation $I^+_n(\lambda)- I^+_n(0)$ by a factor $S$ 
such that the lower-bound approximation $I^-_n(\lambda) - I^-_n(0)$ and the \textit{shifted} upper-bound approximation $I^{+}_n(\lambda) - I^+_n(0)+ S$ are equal at $\lambda^*$.
The shift is obtained by solving the equation
\begin{equation}
I^+_n(\lambda^*) - I^+_n(0) + S = I^-_n(\lambda^*) - I^-_n(0),
\end{equation}
for $S$.
An empirically-motivated and simple choice for $\lambda^*$ is to take $\lambda^* = 1/2$, i.e., to connect the ``$+$'' and ``$-$'' approximations at the point where the reaction is half burned.
%%%%%%%CHECK THE CONDITION BELOW is CORRECT
%%%Another choice is to solve the equation $I^-_n(\lambda^*) -I^+_n(\lambda^*) = 0$ for $\lambda^*$, although this approach relies on numerical root finding which increases the complexity of the method.
%a complete solution that is continuous in $\lambda$ is then given by the piecewise solution
%\begin{equation}
%     \label{eq:Ifull}
%  I_n(\lambda) \approx \left\{
%     \begin{array}{lc}
%		I^-_n(\lambda), &  \lambda < 1/2 \\
%       I^+_n(\lambda) + I^-_n(1/2) -I^+_n(1/2) , &  \lambda \geq 1/2			 
%     \end{array}\;.
%   \right.
%\end{equation}
With $\lambda^* = 1/2$, a complete solution to the Arrenhius-Semenov ODE that is continuous in $\lambda$ is given by 
%Applying the boundary condition $\lambda(t =0) = 0$ to the lower-bound solution yields a complete solution for the Semenov ODE
\begin{equation}
     \label{eq:Ifull}
  k t = I_n(\lambda) - I_n(0) \approx \left\{
     \begin{array}{lc}
		I^-_n(\lambda)- I^-_n(0), &  \lambda < 1/2 \\
       I^+_n(\lambda) - I^+_n(1/2)  + I^-_n(1/2) -I^-_n(0) , &  \lambda \geq 1/2			 
     \end{array}\;.
   \right.
\end{equation}
Equation~(\ref{eq:Ifull}) is the primary result in this report.

It is important to note that the analytical approximation in Eq.~(\ref{eq:Ifull}) is not a smooth function and contains a derivative discontinuity at $t(\lambda = 1/2)$.
This lack of smoothness will not be significant 
in situations in which $\lambda(t)$ is the desired quantity, but it may limit the applicability of the
present approach when $\tfrac{d \lambda}{dt}$ is explicitly needed.
%Typically, this derivative discontinuity will be small and the magnitude o
%will decrease with increasing series expansion order.
We have found that, under typically physical conditions, the derivative discontinuity is small 
in comparison to the magnitude of the derivative itself and, as expected, the discontinuity magnitude
%%\begin{equation}
%%\mathcal{DM} = \big|\partial_t I^-_n(1/2)- \partial_t I^+_n(1/2)\big|
%%\end{equation}
decreases with increasing series expansion order. 

\subsection{Approximate Solutions for $n \in \mathbb{N}$ \label{sec:approxsub}}

In the limit $n \to j+1$ with $j \in \mathbb{N}$ (where $n$ is the reaction order), the functions $I^{(j+)}_n(\lambda)$ given in Eqs.~(\ref{eq:intinit})-(\ref{eq:intfinal}) are vertically asymptotic, and  therefore, before applying an initial value $\lambda(t = 0) = 0$, Eq.~(\ref{eq:fullpos}) is divergent for $n \in \mathbb{N}$ if $n \leq \alpha$.
This can be seen in the general expression for these functions:
\begin{equation}
I^{(j+)}_n (\lambda) = \frac{(1-\lambda)^{j+1-n}}{n-j+1} C^{(j+)}_n, 
\end{equation}
where $C^{(j+)}_n$ is a nonzero constant. 
However, when the $j$th order term in the full solution $f^{(j+)}_n(\lambda)$ is constructed that includes the initial value:
\begin{equation}
f^{(j+)}_n(\lambda) = I^{(j+)}_n(\lambda) - I^{(j+)}_n(0),
\end{equation}
we find that 
\begin{equation}
\lim_{n\to j+1} f^{(j+)}_n(\lambda)  = - C^{(j+)}_n \ln \left(1-\lambda\right).
\end{equation}
This shows that the approximate solution does not have a vertical asymptote with respect to variation of $n$.

\section{Exact Solutions for $n \in \mathbb{N}$ \label{sec:exact}}

%%%%%NEED TO CHECK BEFORE PUBLICATION...Do you need both steps
In this section, we derive exact analytical solutions for the RHS of Eq.~(\ref{eq:separable}) for $n \leq 3 : n \in \mathbb{N}$.
%The presentation of these results contradicts Menikoff's statement that there is no analytical solution for $n \neq 1$ \cite{Menikoff2017}.
%In general, the technique used to evaluate $I_n$ for a specific value of $n$ 
%relies on some combination of $u$-substitution, integration by parts, and partial fraction decomposition.
%Below, we state the results for $n \leq 3 : n \in \mathbb{N}$. 
%In the Appendix, we illustrate the full algebra and calculus for the $n = 0$ and $n = 2$ cases.
In all cases the initial value is $\lambda(t = 0) = 0$.

\subsection{$n = 0$}
%We include the $n =0$ solution here for completeness. 
For $n = 0$, the problem reduces to evaluating the integral
\begin{equation}
I_0 = \int \exp\left[\frac{T_a}{T_0 + \mathcal{Q} / C_v \lambda}\right] d \lambda.
\end{equation}
With $u = T_0 + \mathcal{Q} / C_v \lambda$, the integral takes the form
\begin{equation}
I_0 =\frac{C_v}{\mathcal{Q}} \int \exp\left[\frac{T_a}{u}\right] d u.
\end{equation}
Now, let $v = u/T_a = 1/h(\lambda)$, and
\begin{equation}
I_0 = \frac{C_v T_a}{\mathcal{Q}} \int \exp\left[\frac{1}{v}\right] d v,
\end{equation}
which is a known integral form resulting in
\begin{equation}
I_0 = \frac{C_v T_a}{\mathcal{Q}} \Bigg[v \exp\left[\frac{1}{v}\right]- \text{Ei}\left[\frac{1}{v}\right] \Bigg]
\end{equation}
where $\text{Ei}$ is the exponential integral.
After substitution for $v$ we obtain
\begin{equation}
I_0 = \frac{C_v}{\mathcal{Q}} \Bigg[ T(\lambda) \exp\Big[h(\lambda)\Big] - T_a \text{Ei}\Big[h(\lambda)\Big] \Bigg].
\end{equation}
The final solution for $n=0$ is
%%%\begin{equation}
%%%\begin{aligned}
%%%k t &= \text{Ei}\left[h(\lambda)+\frac{T_a}{T_1}\right] - \exp\left[\frac{T_a}{T_1}\right]\text{Ei}\Big[h(\lambda)\Big] - \text{Ei}\left[\frac{T_a}{T_0}\right]+ \exp\left[\frac{T_a}{T_1}\right]\text{Ei}\Big[h(0)\Big],
%%%\end{aligned}
%%%\end{equation}
%%%\begin{equation}
%%%h(\lambda) = \frac{T_a}{T(\lambda)}- \frac{T_a}{T_1}.
%%%\end{equation}
\begin{equation}
\begin{aligned}
\label{eq:I0}
k t &= \left(\frac{C_v}{\mathcal{Q}}\right)  \Bigg[ T(\lambda)\exp\Big[h(\lambda)\Big] - T_a\text{Ei}\Big[h(\lambda)\Big] -  T_0 \exp\left[\frac{T_a}{T_0}\right] + T_a\text{Ei}\left[\frac{T_a}{T_0}\right] \Bigg],
\end{aligned}
\end{equation}
with $h(\lambda)$ defined in Eq.~(\ref{eq:h}) above.

\subsection{$n = 1$}
An exact analytical solution for the $n = 1$ case is known (see Menikoff's derivation in Section 3 of Ref.~\citenum{Menikoff2017}).
Solving $I_1$ gives
%%%\begin{equation}
%%%\begin{aligned}
%%%k t &= \text{Ei}\left[h(\lambda)+\frac{T_a}{T_1}\right] - \exp\left[\frac{T_a}{T_1}\right]\text{Ei}\Big[h(\lambda)\Big] - \text{Ei}\left[\frac{T_a}{T_0}\right]+ \exp\left[\frac{T_a}{T_1}\right]\text{Ei}\Big[h(0)\Big],
%%%\end{aligned}
%%%\end{equation}
%%%\begin{equation}
%%%h(\lambda) = \frac{T_a}{T(\lambda)}- \frac{T_a}{T_1}.
%%%\end{equation}
\begin{equation}
\begin{aligned}
\label{eq:I1}
k t &= \text{Ei}\Big[h(\lambda)\Big] - \exp\left[\frac{T_a}{T_1}\right]\text{Ei}\Big[g(\lambda)\Big] - \text{Ei}\left[\frac{T_a}{T_0}\right]+ \exp\left[\frac{T_a}{T_1}\right]\text{Ei}\Big[g(0)\Big],
\end{aligned}
\end{equation}
%%%\begin{equation}
%%%\begin{aligned}
%%%k t &= \text{Ei}\left(\frac{T_a}{T_0 + \mathcal{Q}/C_v \lambda}\right) - \exp\left(\frac{T_a}{T_1}\right)\text{Ei}\left(\frac{T_a}{T_0 + \mathcal{Q}/C_v \lambda}-\frac{T_a}{T_1 }\right) \\[1ex]
%%%&\quad - \text{Ei}\left(\frac{T_a}{T_0}\right)+ \exp\left(\frac{T_a}{T_1}\right)\text{Ei}\left(\frac{T_a}{T_0 } -\frac{T_a}{T_1 } \right),
%%%\end{aligned}
%%%\end{equation}
with 
\begin{equation}
g(\lambda) = h(\lambda)- \frac{T_a}{T_1}.
\end{equation}
%where $h(\lambda)$ is defined in Eq.~(\ref{eq:h}).
%A step-by-step solution leading to Eq.~(\ref{eq:I1}) can be found in the Appendix.

\subsection{$n = 2$}
For $n = 2$, the integral to evaluate in order to construct the exact solution is
\begin{equation}
I_2 = \int (1-\lambda)^{-2} \exp\left[\frac{T_a}{T_0 + \mathcal{Q} / C_v \lambda}\right] d \lambda.
\end{equation}
Let $u = \tfrac{T_0 + \mathcal{Q} / C_v \lambda}{T_a} = 1/h(\lambda)$. 
%which implies that $\lambda = (T_a u - T_0) C_v / \mathcal{Q}$,
%Then $\tfrac{d u }{d \lambda} = \tfrac{\mathcal{Q}}{C_v T_a}$
The integral now takes the form
\begin{equation}
I_2 =\frac{C_v T_a}{\mathcal{Q}} \int \exp\left[\frac{1}{u}\right] \left(1+ \frac{C_v T_0}{\mathcal{Q}} - \frac{C_v T_a}{\mathcal{Q}} u\right)^{-2} d u.
\end{equation}
Now, let $v = 1/u = h(\lambda)$, then,
% $\tfrac{d v }{d u} = -\tfrac{1}{u^2}$ 
%and
\begin{equation}
I_2 = -\frac{C_v T_a}{\mathcal{Q}} \int \exp\left[v\right] \Bigg(\frac{C_v T_1}{\mathcal{Q}} v - \frac{C_v T_a}{\mathcal{Q}}\Bigg)^{-2} d v.
\end{equation}
Integration by parts and algebraic manipulation gives
\begin{equation}
I_2 = \frac{\mathcal{Q} T_a}{C_v} \Bigg[ \frac{\exp\left[v\right]}{\left(T_1^2 v - T_1 T_a\right)} - \int \frac{\exp\left[v\right]}{\left(T_1^2 v - T_1 T_a\right)} dv \Bigg].
\end{equation}
Finally, let $w = v - T_a/T_1  = g(\lambda)$ 
%such that $\tfrac{d w }{d v} = 1$
in the integral in previous equation, leading to
\begin{equation}
I_2 = \frac{\mathcal{Q} T_a}{C_v} \Bigg[ \frac{\exp\left[v\right]}{\left(T_1^2 v - T_1 T_a\right)} - \frac{1}{T_1^2} \exp\left[\frac{T_a}{T_1}\right] \int \frac{\exp\left[w\right]}{w} dw \Bigg].
\end{equation}
The integral in the second term is the exponential integral:
\begin{equation}
I_2 = \frac{\mathcal{Q} T_a}{C_v} \Bigg[ \frac{\exp\left[v\right]}{\left(T_1^2 v - T_1 T_a\right)} - \frac{1}{T_1^2} \exp\left[\frac{T_a}{T_1}\right] \text{Ei}\left[w\right] \Bigg].
\end{equation}
After substitution for $w$ and $v$ we obtain
\begin{equation}
I_2 = \frac{T(\lambda)}{(1-\lambda) T_1}  \exp\Big[h(\lambda)\Big] 
- \frac{\mathcal{Q} T_a}{C_v T_1^2}  \exp\left[\frac{T_a}{T_1}\right]\text{Ei}\Big[g(\lambda)\Big].
\end{equation}
The final solution for $n=2$ is 
\begin{equation}
\begin{aligned}
\label{eq:I2}
k t &= \frac{T(\lambda)}{T_1 (1-\lambda)}  \exp\Big[h(\lambda)\Big] 
- \frac{\mathcal{Q} T_a}{C_v T_1^2}  \exp\left[\frac{T_a}{T_1}\right]\text{Ei}\Big[g(\lambda)\Big]\\[1ex]
&\quad -\frac{T_0}{T_1}  \exp\left[\frac{T_a}{T_0}\right]
+ \frac{\mathcal{Q} T_a}{C_v T_1^2} \exp\left[\frac{T_a}{T_1}\right]\text{Ei}\Big[g(0)\Big].
\end{aligned}
\end{equation}

\subsection{$n = 3$}
For $n = 3$, the solution is derived using similar techniques to the cases described above.
% with the inclusion 
%of partial fraction decomposition.
After some algebraic manipulation, the solution can be expressed as
\begin{equation}
\begin{aligned}
k t &= \frac{\mathcal{Q}^2 T_a}{2 C_v^2 T_1^3 g(\lambda)}\left(2 + \frac{T_a}{T_1} + \frac{T_a}{T_1 g(\lambda)} \right) \exp\Big[h(\lambda)\Big] - \frac{\mathcal{Q}^2  T_a\left(T_a + 2 T_1\right) 
}{2 C_v^2 T_1^4}  \exp\left[\frac{T_a}{T_1}\right]\text{Ei}\Big[g(\lambda)\Big]\\[1ex]
&\quad -\frac{\mathcal{Q}^2 T_a}{2 C_v^2 T_1^3 g(0)}\left(2 + \frac{T_a}{T_1} + \frac{T_a}{T_1 g(0)} \right) \exp\left[\frac{T_a}{T_0 }\right] +\frac{\mathcal{Q}^2  T_a\left(T_a + 2 T_1\right) 
}{2 C_v^2 T_1^4}  \exp\left[\frac{T_a}{T_1}\right]\text{Ei}\Big[g(0)\Big].
\end{aligned}
\end{equation}

%%%\subsection{$n > 3$}
%%%The general solution strategies mentioned in this Section can also be used to evaluate $I_n$ for reaction orders $n\geq 4:n \in \mathbb{N}$. \CR{Is this true}

%%%%%%%%%%%%%%%%%%%%%%%%%%%
%%%%%\begin{figure}[]
%%%%%%\includegraphics[width = \columnwidth,clip]{./Figs/probdenstwotime.pdf}
%%%%%\includegraphics[width = 8.5cm,clip]{./Figs/probdensfluctwotime.pdf}
%%%%%\caption{\label{fig:probdensfluctwotime}
%%%%%Two-time once-restricted upside (a) and downside (b) probability density for $\rho_0 = \rho^{(\text{ss})}$
%%%%%with threshold energy $E^\ddag = \left\langle E\right\rangle$
%%%%%at different times $t'<t$ shown in the legend of (a).
%%%%%The upside/downside constraint is imposed at $t = 0.1$.
%%%%%The dashed curves correspond to the respective densities at the $t' = t$ limit (equivalent densities are shown in Fig.~\ref{fig:probdensfluc}).
%%%%%}
%%%%%\end{figure}
%%%%%%%%%%%%%%%%%%%%%%%%%%%

\section{Results\label{sec:results}}

%%%%%%%%%%%%%%%%%%%%%%
\begin{figure}[b]
\includegraphics[width = 10.0cm,clip]{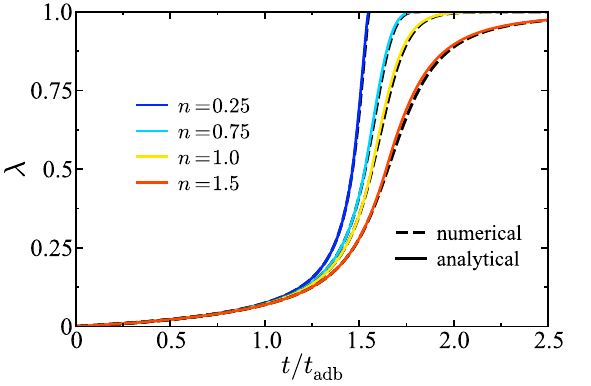}
\caption{\label{fig:Fig1}
Reaction progress variable $\lambda$ as a function of time $t$ (units of $t_\text{adb}$) 
for varying reaction orders $n$ shown in the legend. The system parameters are $T_0 = 800\,\text{K}$, $T_1 = 4000\,\text{K}$, and $T_a = 6000\,\text{K}$. 
The dashed black curves are numerical solutions and the solid curves are the results given by the analytical approximation in Eq.~(\ref{eq:Ifull}) with expansion order $\alpha = 3$.
}
\end{figure}
%%%%%%%%%%%%%%%%%%%%%%

Figure~\ref{fig:Fig1} illustrates a comparison 
between the solution given by the approximation in Eq.~(\ref{eq:Ifull}) and 
the corresponding solution generated numerically for various reaction orders.
In all cases we define time $t$
in units of the adiabatic induction time \cite{Menikoff2017}
\begin{equation}
t_\text{adb} = \frac{C_v T^2_0}{k \mathcal{Q} T_a} \exp\left[\frac{T_a}{T_0}\right],
\end{equation}
and, therefore, the frequency factor $k$ does not enter into the presented results.
All numerical solutions are obtained using the \texttt{Mathematica} program
\texttt{NDSolve} function.
The different color curves correspond to different reaction orders.
The system is characterized by three temperatures: $T_0$, $T_1$, and $T_a$.
In this Figure, the value of these parameters are $T_0 = 800\,\text{K}$, $T_1 = 4000\,\text{K}$, and $T_a = 6000\,\text{K}$
which gives rise to the three stages of ignition observed in a system with large activation temperature  $T_a \gg \tfrac{C_v T_0^2}{\mathcal{Q}}$ :
induction, thermal runway,
%characterized by large $\tfrac{d \lambda}{dt}$
and burn-out.
Over each stage of the burn process, the analytical approximation agrees with the numerical solutions.
The accuracy of approximation decreases as $\lambda$ approaches $1/2$.
The cause of this accuracy decrease is that we expand the upper-bound solution about $\lambda = 1$ and the lower-bond solution 
about $\lambda = 0$, and then connect those solutions at $\lambda = 1/2$. This means that in the complete piecewise solution, $\lambda = 1/2$ is the point furthest from the respective expansion point for each solution branch, and therefore each series generates the greatest error at this point.
In general, the accuracy of the analytical approximation increases as $n \to 0$ and decreases as $n$ is increased.
In order to illustrate the effectiveness of present approach for relatively small expansion order, we have used $\alpha = 3$
in all analytical calculations.
For $n \gg 1$, the 
analytical approximation can also lead to accurate solutions by increasing
the expansion order as illustrated later.
%%%%%\CR{error figure. compare with series expansion of the integral point}

%%%%%%%%%%%%%%%%%%%%%%
\begin{figure}[]
\includegraphics[width = 10.0cm,clip]{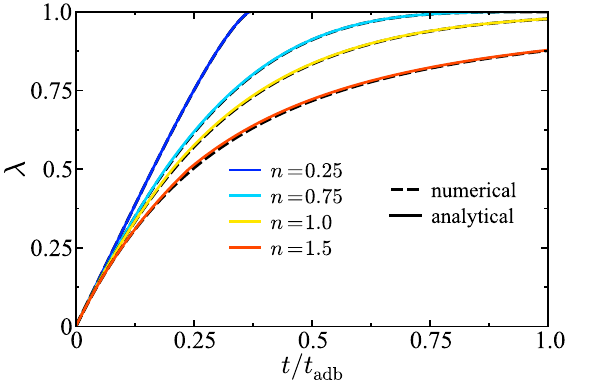}
\caption{\label{fig:Fig2}
Reaction progress variable $\lambda$ as a function of time $t$ (units of $t_\text{adb}$) 
for varying reaction orders $n$ shown in the legend. The system parameters are $T_0 = 3500\,\text{K}$, $T_1 = 4000\,\text{K}$, and $T_a = 8000\,\text{K}$. 
The dashed black curves are numerical solutions and the solid curves are the results given by the analytical approximation in Eq.~(\ref{eq:Ifull}) with expansion order $\alpha = 3$.
}
\end{figure}
%%%%%%%%%%%%%%%%%%%%%%

A comparison between numerical and analytical solutions 
for a constant volume burn with small energy release $\mathcal{Q} \ll \tfrac{C_v T_0^2}{T_a}$ is shown in Fig.~\ref{fig:Fig2} for various reaction orders. 
The system parameters in this Figure are $T_0 = 3500\,\text{K}$, $T_1 = 4000\,\text{K}$, and $T_a = 8000\,\text{K}$.
%which results in exponential-like approach to the burnout stage.
As before, the numerical and analytical results are in agreement,
illustrating that the presented approximation is robust 
with respect to system parameters.

%%%%%%%%%%%%%%%%%%%%%%
\begin{figure}[]
\includegraphics[width = 10.0cm,clip]{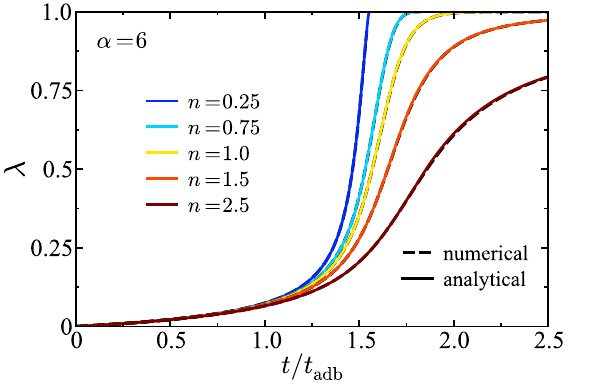}
\caption{\label{fig:Fig3}
Progress variable as a function of time for the same system parameters as Fig.~\ref{fig:Fig1}
with the analytical series expansion order increased to $\alpha = 6$. The different color curves correspond to 
different reaction orders shown in the legend.
}
\end{figure}
%%%%%%%%%%%%%%%%%%%%%%

For simplicity, we have only explicitly shown the algebra for analytical expansions up to third order. Greater solution accuracy, however, can be obtained by increasing
$\alpha$.
Shown in Fig.~\ref{fig:Fig3} is a comparison between analytical and numerical solutions for the same 
system parameters used in Fig.~\ref{fig:Fig1} with expansion order increased to $\alpha = 6$.
The increased value of $\alpha$ results in excellent agreement between solutions for all values of $n$ shown. 
In fact, for small $n$, the analytical result is essentially indistinguishable from the numerical solution at the shown level of visual fidelity.
This illustrates that the analytical method developed here approaches the exact solution 
in a relatively small number of expansion terms. 
Note that we have added a curve for $n= 2.5$ in this plot (cf. Fig.~\ref{fig:Fig1})
% with respect to Fig.~\ref{fig:Fig1}. 
in order to illustrate that accurate solutions for relatively large reaction orders can be constructed
by increasing $\alpha$.

%%%The in $n = $ illustrates that simply Taylor expanding the total integrand in for 
%%%leads to significantly more
%%%\begin{equation}
%%%\int (1-\lambda)^{-n} \exp\left[\frac{T_a}{T_0 + (T_1-T_0) \lambda}\right] d \lambda \approx \int d \lambda \sum^{\alpha'}_{m = 0}
%%%\end{equation}
%%%This illustrates that while a , the piecewise and expansion
%%%leads to 
%%%Illustrating that in a few relatively simple terms the solution to can be well approximated

\section{Acknowledgments}
We thank Josh Coe for insightful discussions.
This work was supported by Los Alamos National Laboratory (LANL) Directed Research and Development funds (LDRD).
\end{document}